\documentclass{appolb}
\usepackage{graphicx}
\usepackage{amssymb}
\usepackage{hyperref}
\usepackage{cleveref}

\begin{document}
\title{Beyond the Standard Model physics in the far-forward region of the Large Hadron Collider%
\thanks{Presented at XXIX Cracow Epiphany Conference on Physics at the Electron-Ion Collider and Future Facilities,
Cracow, Poland, January 16--19, 2023}%
}
\author{Sebastian Trojanowski\footnote{\href{Sebastian.Trojanowski@ncbj.gov.pl}{Sebastian.Trojanowski@ncbj.gov.pl}},
\address{Astrocent, Nicolaus Copernicus Astronomical Center Polish Academy of Sciences, ul. Bartycka 18, 00-716 Warsaw, Poland}
\address{National Centre for Nuclear Research, Pasteura 7, 02-093 Warsaw, Poland}}

\maketitle
\begin{abstract}
A new physics program has been initiated as part of the ongoing LHC physics run in the far-forward region, where dedicated FASER and SND@LHC experiments are currently taking data. We discuss the possible discovery prospects of this program in the search for signatures of beyond the Standard Model physics. We focus on both the present period and the proposed future Forward Physics Facility (FPF) that will operate in the high luminosity LHC era.
\end{abstract}
  
\section{Introduction}

High-energy proton collisions at the Large Hadron Collider (LHC) have, without a doubt, brought us closer than ever to laboratory testing our understanding of very early periods in the evolution of the Universe and revealing the hidden nature of fundamental interactions. Besides the long-awaited discovery of the Higgs boson~\cite{ATLAS:2012yve,CMS:2012qbp}, it has also been an excellent tool for a variety of other studies complementing a theoretical picture of the Standard Model (SM), including measurements in its forward region~\cite{LHCForwardPhysicsWorkingGroup:2016ote}. However, the LHC experiments still need to achieve their maximum sensitivity. A large amount of data will be collected during Run 3 and in the future era of High-Luminosity LHC (HL-LHC). In fact, up-to-date, the integrated luminosity delivered to the ATLAS and CMS experiments corresponds to only about $7.5\%$ of the total luminosity expected at the end of the HL-LHC period~\cite{Gianotti_2023}.

Given the expected long future operation of the LHC and the yet undetermined road ahead for next-generation colliders, it is also time to exploit the full discovery potential of the current machine. The far-forward region of the LHC has proven to be particularly fitting for such explorations. In particular, the new FASER~\cite{FASER:2018ceo,FASER:2018bac,FASER:2019dxq,FASER:2020gpr} and SND@LHC~\cite{SHiP:2020sos,SNDLHC:2022ihg} experiments have recently started measuring high-energy neutrinos produced in $pp$ collisions at the LHC. While this has long been discussed theoretically~\cite{DeRujula:1984pg,Winter:1990ry,DeRujula:1992sn,Vannucci:1993ud,Park:2011gh,Feng:2017uoz}, a proof-of-principle experimental analysis by the FASER collaboration occurred only after the previous LHC Run 2~\cite{FASER:2021mtu}. These capabilities have recently been confirmed using initial Run 3 data to report the first-ever observation of collider neutrinos with a statistical significance of $16$ standard deviations~\cite{FASER:2023zcr}. By the end of the current operation, the ongoing experiments will measure $\mathcal{O}(10^4)$ high-energy neutrino events, including a few tens of tau neutrinos. Significantly, this relies on relatively small detectors with the active volume size of $\mathcal{O}(0.1\,\textrm{m}^3)$. They allow detailed studies of neutrino interaction events at TeV energies at the precision level unattainable at large-scale neutrino facilities.

Measurements in the proposed Forward Physics Facility (FPF) to operate during the HL-LHC era would lead to increased event statistics of additional two orders of magnitude up to $\mathcal{O}(10^6)$ events. The FPF studies would have broad implications for our understanding of proton and nuclear structures and strong interactions, both via measuring the details of neutrino scatterings at different target nuclei and indirectly drawing conclusions about the forward production of parent mesons in $pp$ collisions at the LHC~\cite{Anchordoqui:2021ghd,Feng:2022inv}.

\begin{figure}[t]
\centerline{%
\includegraphics[width=8.5cm]{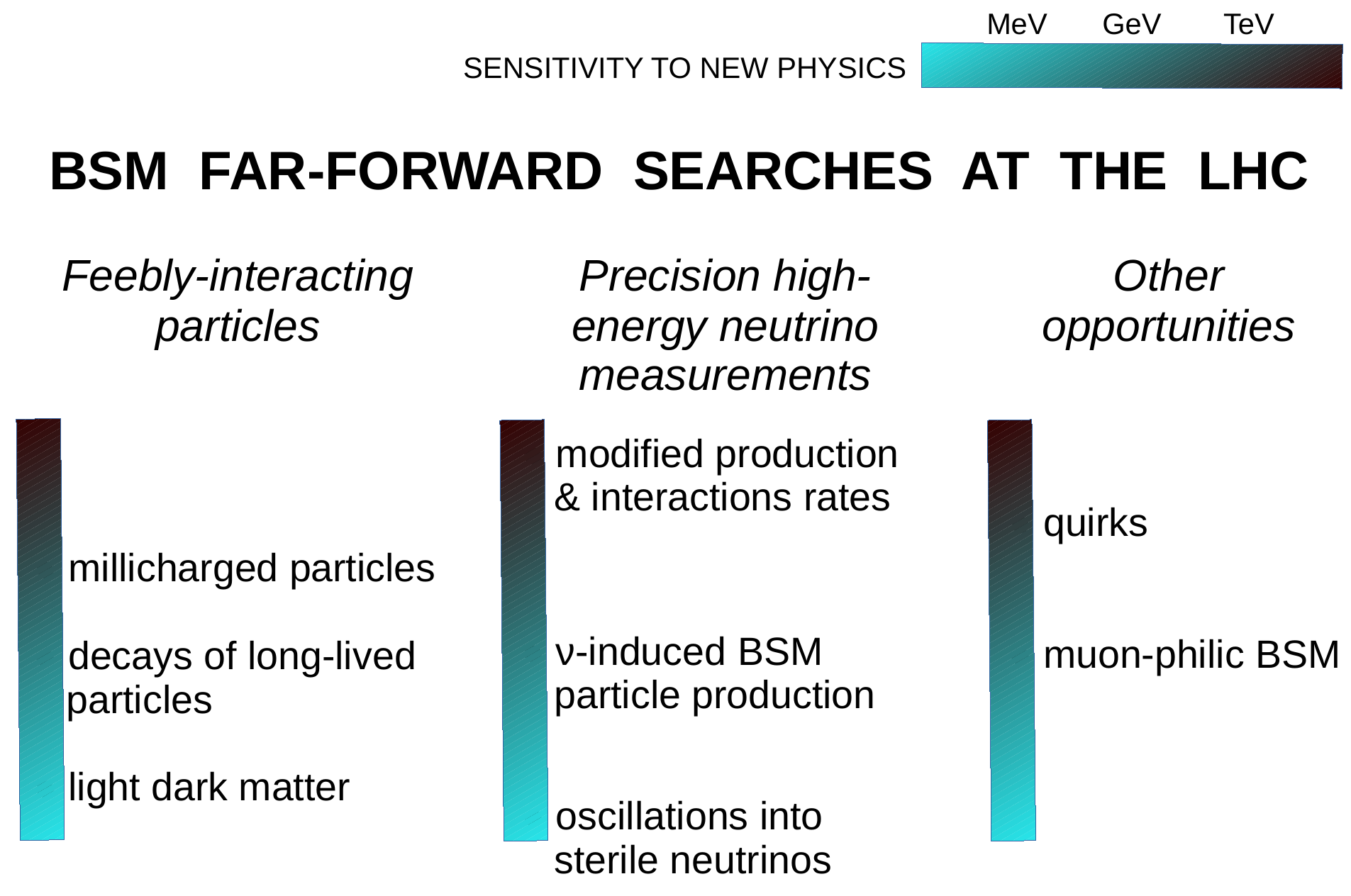}}
\caption{A schematic view of the ongoing and proposed far-forward BSM physics searches at the LHC.}
\label{Fig:schematic}
\end{figure}

A broad experimental program in the LHC's far-forward region also allows for studying physics beyond the Standard Model (BSM), cf. \cref{Fig:schematic} for a schematic view. It includes searches for light new physics species with a mass typically up to tens of GeV detected via their decays, scatterings, or ionization. As we illustrate below, even heavier BSM particles can be studied in specific scenarios, while indirect probes of multi-TeV new physics are also available. In addition, the excellent capabilities of compact far-forward neutrino detectors introduce further opportunities to probe possible tiny deviations from pure SM interactions of neutrinos at energies up to a few TeV scale. BSM physics can also manifest itself in modified neutrino production rates and via oscillations. Below, we provide an overview of the BSM physics opportunities related to far-forward searches at the LHC, which summarizes a much more comprehensive discussion in Ref.~\cite{Feng:2022inv} and mentions some of the most recent results. 

\section{Far-forward experimental program at the LHC}

We first briefly recap experimental aspects of the ongoing and proposed future research program in the far-forward region of the LHC. By far-forward experiments, we mean here detectors placed along the beam collision axis at substantial distances from the $pp$ Interaction Point (IP) such that they remain well-shielded from most SM backgrounds related to proton interactions at the LHC. This shielding requires placing the detectors outside the main LHC tunnel after it curves away from the beam collision axis.

Two such experiments have been installed to take data during the ongoing LHC Run 3: FASER~\cite{FASER:2018ceo,FASER:2018bac} and SND@LHC~\cite{SHiP:2020sos,SNDLHC:2022ihg}. They are placed symmetrically on both sides of the ATLAS IP at a distance about $L\simeq 480~\textrm{m}$ away from it. The FASER detector is devoted to searching for postulated BSM light long-lived particles (LLPs) that can be abundantly produced in the forward region of the LHC and, subsequently, decay visibly in an empty detector volume, as initially proposed in Refs~\cite{Feng:2017uoz,Feng:2017vli,Kling:2018wct,Feng:2018pew}, cf. also Ref.~\cite{FASER:2018eoc} for further discussion. The charged particles (e.g., $e^+e^-$ pairs) produced in such decays travel through the FASER spectrometer and calorimeter for their measurement and identification. The FASER$\nu$ emulsion subdetector~\cite{FASER:2019dxq,FASER:2020gpr} studies interactions of high-energy neutrinos produced at the LHC. The emulsion detector technology and additional electronic components are also used in the SND@LHC experiment~\cite{SHiP:2020sos,SNDLHC:2022ihg}. The FASER detector is placed along the beam collision axis covering the pseudorapidity region of $\eta\gtrsim 9.1$ in the neutrino measurements, while the SND@LHC experiment is slightly displaced, and its measurements correspond to $7.2\lesssim \eta\lesssim 8.4$.

A significant extension of this experimental program has been proposed for the HL-LHC era. A purpose-built Forward Physics Facility~\cite{Anchordoqui:2021ghd,Feng:2022inv} could host a suite of detectors, including successors of the current experiments: FASER2, FASER$\nu$2, and Advanced SND@LHC that will significantly improve the sensitivity of the current BSM and SM far-forward physics searches. In addition, so far two other experiments have been proposed to operate in the FPF: liquid-argon time projection chamber (LArTPC) detector, dubbed FLArE, to study neutrino interactions and search for light dark matter (DM) species~\cite{Batell:2021blf} and a milliQan-type detector FORMOSA targeting new postulated millicharged particles (mCPs)~\cite{Foroughi-Abari:2020qar}. The considered location of the new underground facility would allow placing the experiments on the beam collision axis at a distance of $L\gtrsim 620\,\textrm{m}$ away from the ATLAS IP. The detectors would be shielded from the IP by more than $200\,\textrm{m}$ of rock and concrete to ensure radiation safety and a low-background environment. An important source of backgrounds is remnant muons, i.e., muons produced at the IP and not deflected away by the LHC magnets. The ongoing far-forward experiments, however, have already demonstrated excellent capabilities in rejecting muon-induced events~\cite{FASER:2018bac,SNDLHC:2022ihg,BrianFASER}.

\section{Feebly-interacting particles}

Driven by both theoretical appeal and the history of past discoveries in particle physics, the primary focus of the BSM physics program at the LHC has been on heavy new species and high-$p_T$ searches. While this program is continued and expanded, the importance of going beyond this paradigm has also been realized, particularly in connection to searches for light and feebly interacting particles~\cite{Beacham:2019nyx,Alimena:2019zri,Antel:2023hkf}. Such new physics species with the mass $m\lesssim \mathcal{O}(10~\textrm{GeV})$ can preferentially be produced with low transverse momentum $p_T\sim m$ and travel along the beam collision axis, $p_T/p \ll 1$ for $p\sim \textrm{TeV}$. Based on various experimental signatures, the far-forward experiments at the LHC can probe LLP particles with masses up to a few tens of GeV.

Essential for the current LHC Run 3 are searches for very light species with $m\lesssim m_\pi$ that can come from rare pions and other meson's decays. One expects even $\textrm{a few} \times 10^{17}$ pions to be produced in this period of LHC operation~\cite{FASER:2018eoc}. Nearly $2\%$ of such pions with $E_{\pi}\gtrsim 10~\textrm{GeV}$ travel forward towards the FASER experiment, even though the angular size of this detector is very small, of the order of $10^{-8}$ part of the forward hemisphere. Hence, new light species produced in rare decays of pions and other light mesons can form a highly collimated flux of particles going through the far-forward experiments. 

A vanilla example of such a model is a secluded dark photon $A^\prime$ which couples to the SM via kinetic mixing~\cite{Okun:1982xi,Holdom:1985ag}. Light dark photons mostly come from rare pions and eta mesons decays, $(\pi/\eta)\to \gamma A^\prime$. For the kinetic mixing parameter in the range $\epsilon \sim (10^{-6}-10^{-4})$, this yields up to $\mathcal{O}(10^8)$ high-energy $A^\prime$s produced in the forward region of the LHC~\cite{Feng:2017uoz}. Boosted with characteristic LHC energies, they travel sizable distances until they decay into an energetic $e^+e^-$ or $\mu^+\mu^-$ pair. Null results of the search for such decays inside the FASER detector provide the leading bounds on dark photons in this $(m_{A^\prime},\epsilon)$ range based on initial Run 3 data~\cite{BrianFASER}. Probing larger masses of the dark photon in the far-forward region of the LHC relies on other production modes, e.g., proton bremsstrahlung, cf. Refs~\cite{Feng:2017uoz,Blumlein:2013cua,deNiverville:2016rqh,Foroughi-Abari:2021zbm}. FASER(2) sensitivity to light dark vector bosons that arise from gauging various global SM symmetries have been studied in Refs~\cite{FASER:2018eoc,Bauer:2018onh,Bauer:2020itv,Kling:2020iar,Batell:2021snh,Foguel:2022ppx}.

\begin{figure}[t]
\centerline{%
\includegraphics[width=8.5cm]{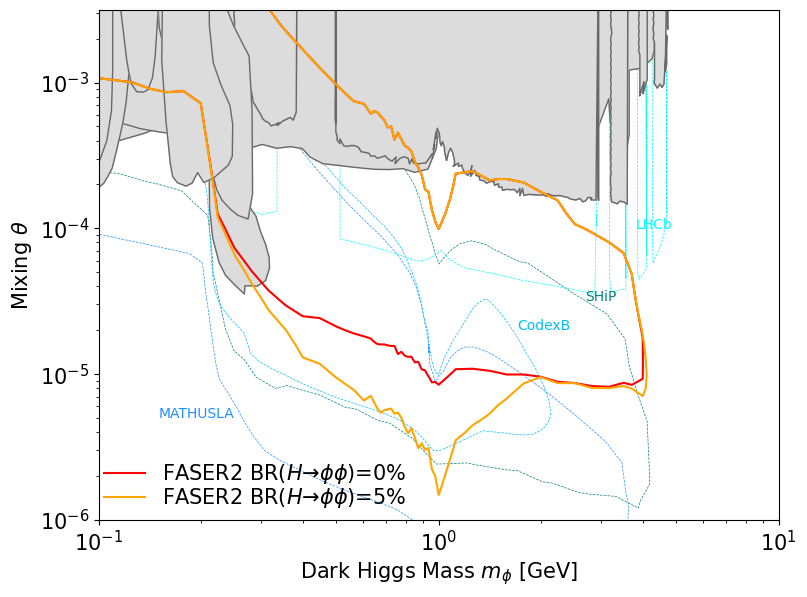}}
\caption{Sensitivity of FASER2 (HL-LHC) in the search for dark Higgs bosons. The plot obtained with \texttt{FORESEE}~\cite{Kling:2021fwx}.}
\label{Fig:darkHiggs}
\end{figure}

Decays of heavier mesons can also lead to large fluxes of LLPs in the far-forward region of the LHC. A prime example is the search for a dark Higgs boson that couples to the SM via mixing with the SM Higgs boson. The light BSM scalar $\phi$ with Yukawa-like couplings to the SM fermions is dominantly produced in rare decays of $B$ mesons. The latter are abundantly produced in high-energy $pp$ collisions at the LHC, e.g., one expects about $N_B\sim 10^{15}$ of $B$ mesons during the HL-LHC era. The corresponding sensitivity reach of the FASER2 experiment proposed for the FPF is shown in \cref{Fig:darkHiggs} in the $(m_{\phi},\theta)$ plane where $\theta$ is the $\phi$-$h$ mixing angle~\cite{Feng:2017vli,FASER:2018eoc,Kling:2021fwx}. We also present there expected sensitivity reach lines for LHCb after the HL-LHC era~\cite{Craik:2022riw} and other selected experiments proposed to search for LLPs: Codex-b~\cite{Aielli:2022awh}, MATHUSLA~\cite{MATHUSLA:2020uve}, SHiP~\cite{Aberle:2839677}. As shown in the plot, the FASER2 reach can be further enhanced by invisible decays of the SM Higgs bosons into lighter scalars, $h^\ast\to \phi\phi$, with the assumed $5\%$ branching fraction. These can be probed indirectly via additional contributions to $B$ meson decays with off-shell $h^\ast$~\cite{Altmannshofer:2009ma}. Similar far-forward experimental programs at future hadron colliders could also probe direct $h\to \phi\phi$ decays of on-shell SM Higgs bosons~\cite{Kling:2021fwx}. Cosmological implications of light dark scalar searches in FASER and beyond have been studied, i.a., in Refs~\cite{Okada:2019opp,Banerjee:2020kww,Csaki:2020zqz}. FASER(2) sensitivity can also be interpreted in terms of Type-I Two Higgs Doublet Model (2HDM)~\cite{Li:2022zgr}, flavor-philic scalar couplings~\cite{Batell:2018fqo}, or a supersymmetric sgoldstino~\cite{Demidov:2022ijc}. 

Above the $B$ meson mass, the primary production mode for LLPs is via a Drell-Yan process. This can lead to a sizeable number of dark photons with a mass up to tens of GeV that can travel along the beam collision axis, although they remain less collimated than lighter LLPs. Prompt and invisible decays of such $A^\prime$s into a secondary flux of dark matter (DM) particles can have significant phenomenological consequences. In particular, unstable inelastic DM (iDM) species produced this way can be probed via their semi-visible decays inside FASER(2), $\chi_2\to\chi_1 e^+e^-$, leading to the bounds for the DM mass inaccessible in experiments at lower energies~\cite{Berlin:2018jbm}. For smaller masses, large boost factors of decaying iDM particles allow for probing very compressed spectra between $\chi_1$ and $\chi_2$ that produce too soft visible signals at lower energies~\cite{Dienes:2023uve}. See also Refs~\cite{Li:2021rzt,Bertuzzo:2022ozu} for further discussion about probing iDM scenarios in FASER.

Light, sub-GeV DM species can also be probed directly in the FPF by searching for their electron scatterings in the FLArE detector~\cite{Batell:2021blf}. Additional nuclear scattering signatures can also be employed~\cite{Batell:2021aja} to probe, e.g., hadrophilic DM scenarios~\cite{Batell:2021snh}, cf. also Ref.~\cite{Boyarsky:2021moj} for a discussion about the (Advanced) SND@LHC experiment. Neutrino-induced backgrounds need to be suppressed in this search. Given the typical high energy of far-forward neutrinos, $E_\nu\gtrsim 100~\textrm{GeV}$, this becomes more straightforward for small energy depositions in the detector characteristic for BSM interactions mediated by light (sub-GeV) new particles. The electron scattering signature can equally well probe BSM interactions of high-energy neutrinos, e.g., in a dipole or $Z^\prime$ portal model to heavy neutral leptons (HNLs)~\cite{Jodlowski:2020vhr,Ismail:2021dyp,Ovchynnikov:2023wgg,Liu:2023nxi} or for dark vector mediators coupled to the lepton number~\cite{Cheung:2022kjd}. Unstable but extremely long-lived species can also be detected, e.g., via scattering processes $N e\to N e$, where $N$ is a right-handed neutrino~\cite{Jodlowski:2020vhr}.

Postulated light new physics species with a fractional electric charge $Q_\chi$ would leave ionization signatures instead. The LHC in its far-forward region is a factory of such mCPs with a mass up to tens of GeV. The search for them in the FORMOSA experiment, to operate in the FPF, is expected to deliver leading bounds on mCPs in the mass range up to $m\lesssim 100~\textrm{GeV}$ and probe mCPs with electric charge as small as $Q_\chi\lesssim 10^{-3}\,e$~\cite{Foroughi-Abari:2020qar}. Additional constraints on such particles can result from the search for their expected electron scatterings in the FLArE detector~\cite{Kling:2022ykt}.

\begin{figure}[t]
\centerline{%
\includegraphics[width=8.5cm]{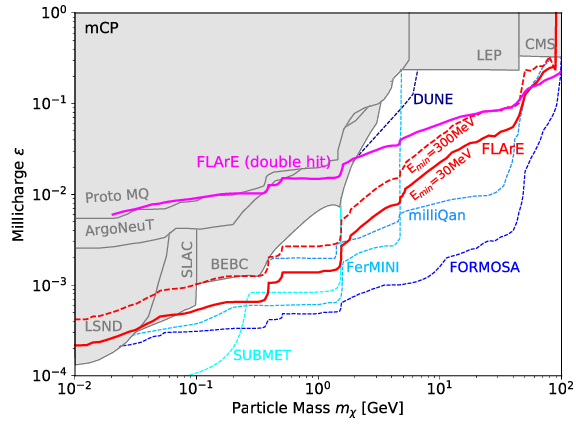}}
\caption{Sensitivity of FORMOSA and FLArE experiments in the search for millicharged particles. From Ref.~\cite{Kling:2022ykt}.}
\label{Fig:mCP}
\end{figure}

Besides the $pp$ collisions, forward-going LLPs can also be produced in other parts of the LHC tunnel and beyond, in the material in front of the detector. For example, the production of axion-like particles (ALPs) coupled to photons in the Primakoff process in high-energy photon interactions determines FASER(2) sensitivity to this scenario~\cite{Feng:2018pew}, cf. also Ref.~\cite{Kling:2022ehv} for a discussion about FASER working as a light shining through a wall experiment. This secondary production of LLPs can also occur close to the detector or inside, leading to enhanced sensitivity in a low LLP lifetime regime~\cite{Jodlowski:2019ycu}, cf. also Refs~\cite{Jodlowski:2020vhr,Bakhti:2020szu,Ansarifard:2021elw} for further discussion including BSM-induced high-energy photon and multi-muon signatures.

Phenomenological studies related to LLP searches in FASER and FPF experiments can be significantly simplified with a dedicated simulation package \texttt{FORESEE}~\cite{Kling:2021fwx}. The tool allows one to run the entire analysis to obtain sensitivity reach plots for a growing library of BSM models and extract valuable data for external simulations, e.g., forward hadron spectra. In \texttt{FORESEE}, detector geometries, positions, and experimental cuts are free to vary by the user, which can be used likewise to study other ideas related to the forward region of the LHC~\cite{Cerci:2021nlb}. The package additionally allows for testing LLP model sensitivity of potential future far-forward experimental programs in hadronic colliders at $27$ and $100~\textrm{TeV}$ energies.

The discovery prospects of FASER(2) in the search for light and feebly interacting particles are broader than the scenarios outlined above. Far-forward experiments are sensitive to popular GeV-scale heavy neutral leptons (HNLs) that couple to SM neutrinos via mass mixing~\cite{Kling:2018wct,Helo:2018qej} or higher-dimensional operators~\cite{DeVries:2020jbs,Cottin:2021lzz}. Similarly, light R-parity violating supersymmetric neutralinos can be probed via their predicted highly-displaced decays~\cite{Dercks:2018eua,Dreiner:2022swd}. The far-forward experiments can also constrain higher dimensional couplings between dark and visible fermions~\cite{Darme:2020ral} and models predicting lepton-number violation~\cite{Araki:2022xqp,Mao:2023zzk}. FASER(2) sensitivity in the search for axion-like particles has also been extensively analyzed for various interaction terms. Besides the ALP-photon coupling mentioned above~\cite{Feng:2018pew}, phenomenological consequences of ALP interaction with SM fermions and gluons~\cite{FASER:2018eoc}, $SU(2)_L$ gauge bosons~\cite{Kling:2020mch}, or flavor-specific couplings to up-type quarks~\cite{Carmona:2021seb} have been studied. A comprehensive review of these and other LLP searches at the FPF can be found in Ref.~\cite{Feng:2022inv}.

\section{Neutrino BSM interactions}

As mentioned above, precision high-energy neutrino physics is fundamental to far-forward searches at the LHC. Besides the rich SM experimental program, it can also be used for probing possible neutrino BSM interactions. These can affect neutrino production, propagation, and final detection through scattering processes.

While electron and muon neutrinos are abundantly produced in decays of light mesons, and, therefore, there is little room to probe them via modified production rates in $pp$ collisions at the LHC, this is not the case for tau neutrinos. These come predominantly from $D$ meson and subsequent tau lepton decays, which results in up to three orders of magnitude lower expected flux of $\nu_\tau$s than the dominant $\nu_\mu$ flux~\cite{Kling:2021gos}. If tau neutrinos can be additionally produced, e.g., in rare BSM-induced pion decays, new physics can manifest itself as an excess of $\nu_\tau$ charged current (CC) scattering events in FASER$\nu$2~\cite{Kling:2020iar}, cf. also Refs~\cite{Batell:2021snh,Bahraminasr:2020ssz,Ansarifard:2021dju} for further illustration. This search requires carefully treating far-forward tau neutrino flux uncertainties and potentially exploiting full neutrino scattering events' energy and position information to reduce them in the search for BSM phenomena, cf. Ref.~\cite{Feng:2022inv} and references therein for further discussion about these uncertainties. Forward neutrino production rates can also be modified in the presence of lepton flavor non-universality~\cite{Aloni:2022ebm}.

Neutrino BSM couplings to charged leptons and quarks can, generally, affect both their production and detection process. At the effective field theory (EFT) level, this has been comprehensively studied in Ref.~\cite{Falkowski:2021bkq}. As shown therein, the lack of predicted deviations from SM neutrino interaction rates can lead to FASER$\nu$ bounds up to even a multi-TeV scale of new physics already during the ongoing LHC Run 3. These results can then be translated into specific BSM frameworks, e.g., models predicting heavy lepto-quarks~\cite{Feng:2022inv}, cf. also Refs~\cite{Cheung:2021tmx,Cheung:2023gwm} for further discussion about probing heavy BSM species via neutrino interactions in the FPF.

The presence of high-energy neutrinos in the FPF additionally opens up the possibility of their upscattering into dark species with the mass $m\sim 1\,\textrm{GeV}$. While kinematically inaccessible in low-energy experiments, BSM effects of this type are also challenging to probe in atmospheric neutrino interactions in large-scale neutrino telescopes. These remain insensitive to subtle details of neutrino scattering events that could distinguish new physics contributions. Examples of such studies concern FPF searches for effects of neutrino upscatterings driven by neutrino dipole portal coupling to HNLs~\cite{Jodlowski:2020vhr,Ismail:2021dyp,Cheung:2022oji}. BSM species can also be produced in neutrino CC scatterings leading to anomalous patterns in the transverse momentum distribution of the outgoing charged particles, as illustrated for a popular model with a neutrino portal to DM in Ref.~\cite{Kelly:2021mcd}.

Measuring neutral current (NC) scatterings in the ongoing and future far-forward searches at the LHC remains more challenging, given the significant missing energy in the final state carried out by a final-state neutrino. The corresponding potential for NC event identification and neutrino energy measurement has been studied, however, with full event information available from precise emulsion detectors and by employing deep neural networks~\cite{Ismail:2020yqc}. This analysis allows also for probing non-standard interactions (NSI) of neutrinos by determining the NC/CC ratio. Anomalous NC nuclear or electron scattering rates could further indicate new physics effects related to non-vanishing neutrino electromagnetic moments. In particular, the FPF can deliver leading bounds on charge-radius contributions of this type for electron neutrinos~\cite{MammenAbraham:2023psg}.

On their way to forward experiments at the LHC, neutrinos are negligibly affected by oscillations, given their typical energy and the distance to the detectors. Significant signs of such oscillations could then be indicative of new physics. These could, e.g., manifest as suppression of events in specific energy bins, cf. Ref.~\cite{FASER:2019dxq} where such a $\nu_\mu$ disappearance effect has been discussed for oscillations into sterile neutrinos with $\Delta m_{41}^2\sim 2000\,\textrm{eV}^2$. Further analysis shows that this channel could provide leading bounds down to mixing angles $U_{\mu 4}^2<10^{-2}$ for $\Delta m_{41}^2\sim (100 - 1000)\,\textrm{eV}^2$~\cite{Feng:2022inv}. Neutrino spectrum shape uncertainties must be considered to estimate the impact of BSM oscillations correctly~\cite{Bai:2020ukz}. 

Last but not least, the far-forward LHC neutrino physics program can also be used to study predicted SM phenomena that could not have been better measured before or completely avoided detection up-to-date. A prime example of the former event type is $\nu_\tau$ CC scatterings. Ultimately, the FPF would measure $\mathcal{O}(10^3)$ such interactions, dramatically increasing the available statistics of such events that have ever been available. These statistics could be further enhanced, e.g., in the presence of intrinsic charm contribution to the $\nu_\tau$ flux from excess production of charm mesons in the far-forward region of the LHC~\cite{Maciula:2022lzk}. Finally, extremely rare SM events that require high scattering energy might become available for measurements in the FPF, e.g., $\bar{\nu}_e e^-\to \rho^-\to \pi^-\pi^0$ process~\cite{Brdar:2021hpy}.

\section{Other BSM physics opportunities}

Far-forward searches at the LHC can also directly probe heavy new particles with masses up to $\mathcal{O}(1\,\textrm{TeV})$ if they are abundantly produced along the beam collision axis. A notable example is the proposed FASER(2) search for quirks charged under the SM and a hidden confining gauge group~\cite{Li:2021tsy}. Quirks can be effectively pair-produced in $pp$ collisions at the LHC with a low combined transverse momentum of the pair. They then travel forward and could leave non-trivial and unique signatures in the detectors, complementing quirk searches in experiments operating at higher $p_T$s.

Additional opportunities for precision SM measurements and BSM searches could be related to a large flux of high-energy muons traveling through the far-forward experiments. While strong LHC magnets deflect away most of the muons produced at ATLAS IP, one still expects $\mathcal{O}(10^9)$ of them to reach FASER during LHC Run 3~\cite{FASER:2018ceo,FASER:2019dxq}. This number further grows for the FPF in the HL-LHC era. In particular, characteristic TeV energies of LHC muons could allow for probing light muon-philic species produced in BSM muon interactions in the forward detector~\cite{Muontalk,Ariga:2023fjg}. The full potential of such a possible muon physics program at the LHC remains to be studied in detail.

\section{Summary}

A new experimental program at the LHC has started during LHC Run 3 in its far-forward region with dedicated FASER and SND@LHC experiments. It focuses on precise high-energy neutrino measurements and related QCD studies and searches for new physics. The BSM physics program of the currently running experiments and the future proposed Forward Physics Facility is rich. It is devoted primarily to searches for postulated light feebly interacting particles and probing possible BSM neutrino interactions, although it is not limited to these topics. Various new physics scenarios have already been discussed in this context, and they connect to essential open questions in contemporary particle physics~\cite{Feng:2022inv}. 

Besides the proposed experiments, the FPF could offer further opportunities for new searches and measurements to explore the LHC physics potential in its far-forward region fully. New ideas of this type are constantly discussed in the FPF community, which remains open to contributions from researchers at all stages of their careers. An essential aspect of the FPF studies is an interplay between the SM and BSM physics programs. The latter will significantly benefit from the ongoing and future neutrino measurements and constraints imposed on forward hadron production obtained this way.
\smallskip

\textbf{Acknowledgements} ST would like to thank Brian Batell, Jonathan Feng, and Felix Kling for useful discussions. This work is supported by the grant ``AstroCeNT: Particle Astrophysics Science and Technology Centre'' carried out within the International Research Agendas programme of the Foundation for Polish Science financed by the European Union under the European Regional Development Fund and by the National Science Centre, Poland, research grant No. 2021/42/E/ST2/00031.

\bibliographystyle{utphys}
\bibliography{references} 

\providecommand{\href}[2]{#2}\begingroup\raggedright\begin{thebibliography}{10}

\bibitem{ATLAS:2012yve}
{\bfseries ATLAS} Collaboration, G.~Aad {\em et al.}, ``{Observation of a new
  particle in the search for the Standard Model Higgs boson with the ATLAS
  detector at the LHC},''
  \href{http://dx.doi.org/10.1016/j.physletb.2012.08.020}{{\em Phys. Lett. B}
  {\bfseries 716} (2012) 1--29},
  \href{http://arxiv.org/abs/1207.7214}{{\ttfamily arXiv:1207.7214 [hep-ex]}}.

\bibitem{CMS:2012qbp}
{\bfseries CMS} Collaboration, S.~Chatrchyan {\em et al.}, ``{Observation of a
  New Boson at a Mass of 125 GeV with the CMS Experiment at the LHC},''
  \href{http://dx.doi.org/10.1016/j.physletb.2012.08.021}{{\em Phys. Lett. B}
  {\bfseries 716} (2012) 30--61},
  \href{http://arxiv.org/abs/1207.7235}{{\ttfamily arXiv:1207.7235 [hep-ex]}}.

\bibitem{LHCForwardPhysicsWorkingGroup:2016ote}
{\bfseries LHC Forward Physics Working Group} Collaboration, K.~Akiba {\em et
  al.}, ``{LHC Forward Physics},''
  \href{http://dx.doi.org/10.1088/0954-3899/43/11/110201}{{\em J. Phys. G}
  {\bfseries 43} (2016) 110201},
  \href{http://arxiv.org/abs/1611.05079}{{\ttfamily arXiv:1611.05079
  [hep-ph]}}.

\bibitem{Gianotti_2023}
F.~Gianotti, ``CERN visions and plans.''
  {\href{https://indico.bnl.gov/event/18372/contributions/75206/attachments/47011/79716/CERN-plans.pdf}{talk
  at the P5 Town Hall meeting on the Future of High Energy Physics, April
  2023}}.

\bibitem{FASER:2018ceo}
{\bfseries FASER} Collaboration, A.~Ariga {\em et al.}, ``{Letter of Intent for
  FASER: ForwArd Search ExpeRiment at the LHC},''
  \href{http://arxiv.org/abs/1811.10243}{{\ttfamily arXiv:1811.10243
  [physics.ins-det]}}.

\bibitem{FASER:2018bac}
{\bfseries FASER} Collaboration, A.~Ariga {\em et al.}, ``{Technical Proposal
  for FASER: ForwArd Search ExpeRiment at the LHC},''
  \href{http://arxiv.org/abs/1812.09139}{{\ttfamily arXiv:1812.09139
  [physics.ins-det]}}.

\bibitem{FASER:2019dxq}
{\bfseries FASER} Collaboration, H.~Abreu {\em et al.}, ``{Detecting and
  Studying High-Energy Collider Neutrinos with FASER at the LHC},''
  \href{http://dx.doi.org/10.1140/epjc/s10052-020-7631-5}{{\em Eur. Phys. J. C}
  {\bfseries 80} no.~1, (2020) 61},
  \href{http://arxiv.org/abs/1908.02310}{{\ttfamily arXiv:1908.02310
  [hep-ex]}}.

\bibitem{FASER:2020gpr}
{\bfseries FASER} Collaboration, H.~Abreu {\em et al.}, ``{Technical Proposal:
  FASERnu},'' \href{http://arxiv.org/abs/2001.03073}{{\ttfamily
  arXiv:2001.03073 [physics.ins-det]}}.

\bibitem{SHiP:2020sos}
{\bfseries SHiP} Collaboration, C.~Ahdida {\em et al.}, ``{SND@LHC},''
  \href{http://arxiv.org/abs/2002.08722}{{\ttfamily arXiv:2002.08722
  [physics.ins-det]}}.

\bibitem{SNDLHC:2022ihg}
{\bfseries SND@LHC} Collaboration, G.~Acampora {\em et al.}, ``{SND@LHC: The
  Scattering and Neutrino Detector at the LHC},''
  \href{http://arxiv.org/abs/2210.02784}{{\ttfamily arXiv:2210.02784
  [hep-ex]}}.

\bibitem{DeRujula:1984pg}
A.~De~Rujula and R.~Ruckl,
  \href{http://dx.doi.org/10.5170/CERN-1984-010-V-2.571}{``{Neutrino and muon
  physics in the collider mode of future accelerators},''} in {\em {SSC
  Workshop: Superconducting Super Collider Fixed Target Physics}},
  pp.~571--596.
\newblock 5, 1984.

\bibitem{Winter:1990ry}
K.~Winter, ``{Detection of the tau-neutrino at the LHC},'' in {\em {ECFA Large
  Hadron Collider (LHC) Workshop: Physics and Instrumentation}}, pp.~37--49.
\newblock 1990.

\bibitem{DeRujula:1992sn}
A.~De~Rujula, E.~Fernandez, and J.~J. Gomez-Cadenas, ``{Neutrino fluxes at
  future hadron colliders},''
  \href{http://dx.doi.org/10.1016/0550-3213(93)90427-Q}{{\em Nucl. Phys. B}
  {\bfseries 405} (1993) 80--108}.

\bibitem{Vannucci:1993ud}
F.~Vannucci, ``{Neutrino physics at LHC / SSC},'' in {\em {4th International
  Symposium on Neutrino Telescopes}}.
\newblock 3, 1993.

\bibitem{Park:2011gh}
H.~Park, ``{The estimation of neutrino fluxes produced by proton-proton
  collisions at $\sqrt{s}=14$ TeV of the LHC},''
  \href{http://dx.doi.org/10.1007/JHEP10(2011)092}{{\em JHEP} {\bfseries 10}
  (2011) 092}, \href{http://arxiv.org/abs/1110.1971}{{\ttfamily arXiv:1110.1971
  [hep-ex]}}.

\bibitem{Feng:2017uoz}
J.~L. Feng, I.~Galon, F.~Kling, and S.~Trojanowski, ``{ForwArd Search
  ExpeRiment at the LHC},''
  \href{http://dx.doi.org/10.1103/PhysRevD.97.035001}{{\em Phys. Rev. D}
  {\bfseries 97} no.~3, (2018) 035001},
  \href{http://arxiv.org/abs/1708.09389}{{\ttfamily arXiv:1708.09389
  [hep-ph]}}.

\bibitem{FASER:2021mtu}
{\bfseries FASER} Collaboration, H.~Abreu {\em et al.}, ``{First neutrino
  interaction candidates at the LHC},''
  \href{http://dx.doi.org/10.1103/PhysRevD.104.L091101}{{\em Phys. Rev. D}
  {\bfseries 104} no.~9, (2021) L091101},
  \href{http://arxiv.org/abs/2105.06197}{{\ttfamily arXiv:2105.06197
  [hep-ex]}}.

\bibitem{FASER:2023zcr}
{\bfseries FASER} Collaboration, H.~Abreu {\em et al.}, ``{First Direct
  Observation of Collider Neutrinos with FASER at the LHC},''
  \href{http://arxiv.org/abs/2303.14185}{{\ttfamily arXiv:2303.14185
  [hep-ex]}}.

\bibitem{Anchordoqui:2021ghd}
L.~A. Anchordoqui {\em et al.}, ``{The Forward Physics Facility: Sites,
  experiments, and physics potential},''
  \href{http://dx.doi.org/10.1016/j.physrep.2022.04.004}{{\em Phys. Rept.}
  {\bfseries 968} (2022) 1--50},
  \href{http://arxiv.org/abs/2109.10905}{{\ttfamily arXiv:2109.10905
  [hep-ph]}}.

\bibitem{Feng:2022inv}
J.~L. Feng {\em et al.}, ``{The Forward Physics Facility at the High-Luminosity
  LHC},'' \href{http://dx.doi.org/10.1088/1361-6471/ac865e}{{\em J. Phys. G}
  {\bfseries 50} no.~3, (2023) 030501},
  \href{http://arxiv.org/abs/2203.05090}{{\ttfamily arXiv:2203.05090
  [hep-ex]}}.

\bibitem{Feng:2017vli}
J.~L. Feng, I.~Galon, F.~Kling, and S.~Trojanowski, ``{Dark Higgs bosons at the
  ForwArd Search ExpeRiment},''
  \href{http://dx.doi.org/10.1103/PhysRevD.97.055034}{{\em Phys. Rev. D}
  {\bfseries 97} no.~5, (2018) 055034},
  \href{http://arxiv.org/abs/1710.09387}{{\ttfamily arXiv:1710.09387
  [hep-ph]}}.

\bibitem{Kling:2018wct}
F.~Kling and S.~Trojanowski, ``{Heavy Neutral Leptons at FASER},''
  \href{http://dx.doi.org/10.1103/PhysRevD.97.095016}{{\em Phys. Rev. D}
  {\bfseries 97} no.~9, (2018) 095016},
  \href{http://arxiv.org/abs/1801.08947}{{\ttfamily arXiv:1801.08947
  [hep-ph]}}.

\bibitem{Feng:2018pew}
J.~L. Feng, I.~Galon, F.~Kling, and S.~Trojanowski, ``{Axionlike particles at
  FASER: The LHC as a photon beam dump},''
  \href{http://dx.doi.org/10.1103/PhysRevD.98.055021}{{\em Phys. Rev. D}
  {\bfseries 98} no.~5, (2018) 055021},
  \href{http://arxiv.org/abs/1806.02348}{{\ttfamily arXiv:1806.02348
  [hep-ph]}}.

\bibitem{FASER:2018eoc}
{\bfseries FASER} Collaboration, A.~Ariga {\em et al.},
  ``{FASER\textquoteright{}s physics reach for long-lived particles},''
  \href{http://dx.doi.org/10.1103/PhysRevD.99.095011}{{\em Phys. Rev. D}
  {\bfseries 99} no.~9, (2019) 095011},
  \href{http://arxiv.org/abs/1811.12522}{{\ttfamily arXiv:1811.12522
  [hep-ph]}}.

\bibitem{Batell:2021blf}
B.~Batell, J.~L. Feng, and S.~Trojanowski, ``{Detecting Dark Matter with
  Far-Forward Emulsion and Liquid Argon Detectors at the LHC},''
  \href{http://dx.doi.org/10.1103/PhysRevD.103.075023}{{\em Phys. Rev. D}
  {\bfseries 103} no.~7, (2021) 075023},
  \href{http://arxiv.org/abs/2101.10338}{{\ttfamily arXiv:2101.10338
  [hep-ph]}}.

\bibitem{Foroughi-Abari:2020qar}
S.~Foroughi-Abari, F.~Kling, and Y.-D. Tsai, ``{Looking forward to millicharged
  dark sectors at the LHC},''
  \href{http://dx.doi.org/10.1103/PhysRevD.104.035014}{{\em Phys. Rev. D}
  {\bfseries 104} no.~3, (2021) 035014},
  \href{http://arxiv.org/abs/2010.07941}{{\ttfamily arXiv:2010.07941
  [hep-ph]}}.

\bibitem{BrianFASER}
B.~Petersen, ``First Physics Results from the FASER Experimentfrom the FASER
  Experiment.''
  \href{https://indico.in2p3.fr/event/29681/contributions/122474/attachments/76425/110931/05-BPetersen-v1.pdf}{57th
  Recontres de Moriond}, 2023.

\bibitem{Beacham:2019nyx}
J.~Beacham {\em et al.}, ``{Physics Beyond Colliders at CERN: Beyond the
  Standard Model Working Group Report},''
  \href{http://dx.doi.org/10.1088/1361-6471/ab4cd2}{{\em J. Phys. G} {\bfseries
  47} no.~1, (2020) 010501}, \href{http://arxiv.org/abs/1901.09966}{{\ttfamily
  arXiv:1901.09966 [hep-ex]}}.

\bibitem{Alimena:2019zri}
J.~Alimena {\em et al.}, ``{Searching for long-lived particles beyond the
  Standard Model at the Large Hadron Collider},''
  \href{http://dx.doi.org/10.1088/1361-6471/ab4574}{{\em J. Phys. G} {\bfseries
  47} no.~9, (2020) 090501}, \href{http://arxiv.org/abs/1903.04497}{{\ttfamily
  arXiv:1903.04497 [hep-ex]}}.

\bibitem{Antel:2023hkf}
C.~Antel {\em et al.}, ``{Feebly Interacting Particles: FIPs 2022 workshop
  report},'' \href{http://arxiv.org/abs/2305.01715}{{\ttfamily arXiv:2305.01715
  [hep-ph]}}.

\bibitem{Okun:1982xi}
L.~B. Okun, ``{LIMITS OF ELECTRODYNAMICS: PARAPHOTONS?},'' {\em Sov. Phys.
  JETP} {\bfseries 56} (1982) 502.

\bibitem{Holdom:1985ag}
B.~Holdom, ``{Two U(1)'s and Epsilon Charge Shifts},''
  \href{http://dx.doi.org/10.1016/0370-2693(86)91377-8}{{\em Phys. Lett. B}
  {\bfseries 166} (1986) 196--198}.

\bibitem{Blumlein:2013cua}
J.~Bl\"umlein and J.~Brunner, ``{New Exclusion Limits on Dark Gauge Forces from
  Proton Bremsstrahlung in Beam-Dump Data},''
  \href{http://dx.doi.org/10.1016/j.physletb.2014.02.029}{{\em Phys. Lett. B}
  {\bfseries 731} (2014) 320--326},
  \href{http://arxiv.org/abs/1311.3870}{{\ttfamily arXiv:1311.3870 [hep-ph]}}.

\bibitem{deNiverville:2016rqh}
P.~deNiverville, C.-Y. Chen, M.~Pospelov, and A.~Ritz, ``{Light dark matter in
  neutrino beams: production modelling and scattering signatures at MiniBooNE,
  T2K and SHiP},'' \href{http://dx.doi.org/10.1103/PhysRevD.95.035006}{{\em
  Phys. Rev. D} {\bfseries 95} no.~3, (2017) 035006},
  \href{http://arxiv.org/abs/1609.01770}{{\ttfamily arXiv:1609.01770
  [hep-ph]}}.

\bibitem{Foroughi-Abari:2021zbm}
S.~Foroughi-Abari and A.~Ritz, ``{Dark sector production via proton
  bremsstrahlung},'' \href{http://dx.doi.org/10.1103/PhysRevD.105.095045}{{\em
  Phys. Rev. D} {\bfseries 105} no.~9, (2022) 095045},
  \href{http://arxiv.org/abs/2108.05900}{{\ttfamily arXiv:2108.05900
  [hep-ph]}}.

\bibitem{Bauer:2018onh}
M.~Bauer, P.~Foldenauer, and J.~Jaeckel, ``{Hunting All the Hidden Photons},''
  \href{http://dx.doi.org/10.1007/JHEP07(2018)094}{{\em JHEP} {\bfseries 07}
  (2018) 094}, \href{http://arxiv.org/abs/1803.05466}{{\ttfamily
  arXiv:1803.05466 [hep-ph]}}.

\bibitem{Bauer:2020itv}
M.~Bauer, P.~Foldenauer, and M.~Mosny, ``{Flavor structure of anomaly-free
  hidden photon models},''
  \href{http://dx.doi.org/10.1103/PhysRevD.103.075024}{{\em Phys. Rev. D}
  {\bfseries 103} no.~7, (2021) 075024},
  \href{http://arxiv.org/abs/2011.12973}{{\ttfamily arXiv:2011.12973
  [hep-ph]}}.

\bibitem{Kling:2020iar}
F.~Kling, ``{Probing light gauge bosons in tau neutrino experiments},''
  \href{http://dx.doi.org/10.1103/PhysRevD.102.015007}{{\em Phys. Rev. D}
  {\bfseries 102} no.~1, (2020) 015007},
  \href{http://arxiv.org/abs/2005.03594}{{\ttfamily arXiv:2005.03594
  [hep-ph]}}.

\bibitem{Batell:2021snh}
B.~Batell, J.~L. Feng, M.~Fieg, A.~Ismail, F.~Kling, R.~M. Abraham, and
  S.~Trojanowski, ``{Hadrophilic dark sectors at the Forward Physics
  Facility},'' \href{http://dx.doi.org/10.1103/PhysRevD.105.075001}{{\em Phys.
  Rev. D} {\bfseries 105} no.~7, (2022) 075001},
  \href{http://arxiv.org/abs/2111.10343}{{\ttfamily arXiv:2111.10343
  [hep-ph]}}.

\bibitem{Foguel:2022ppx}
A.~L. Foguel, P.~Reimitz, and R.~Z. Funchal, ``{A robust description of
  hadronic decays in light vector mediator models},''
  \href{http://dx.doi.org/10.1007/JHEP04(2022)119}{{\em JHEP} {\bfseries 04}
  (2022) 119}, \href{http://arxiv.org/abs/2201.01788}{{\ttfamily
  arXiv:2201.01788 [hep-ph]}}.

\bibitem{Kling:2021fwx}
F.~Kling and S.~Trojanowski, ``{Forward experiment sensitivity estimator for
  the LHC and future hadron colliders},''
  \href{http://dx.doi.org/10.1103/PhysRevD.104.035012}{{\em Phys. Rev. D}
  {\bfseries 104} no.~3, (2021) 035012},
  \href{http://arxiv.org/abs/2105.07077}{{\ttfamily arXiv:2105.07077
  [hep-ph]}}.

\bibitem{Craik:2022riw}
D.~Craik, P.~Ilten, D.~Johnson, and M.~Williams, ``{LHCb future dark-sector
  sensitivity projections for Snowmass 2021},'' in {\em {Snowmass 2021}}.
\newblock 3, 2022.
\newblock \href{http://arxiv.org/abs/2203.07048}{{\ttfamily arXiv:2203.07048
  [hep-ph]}}.

\bibitem{Aielli:2022awh}
G.~Aielli {\em et al.}, ``{The Road Ahead for CODEX-b},''
  \href{http://arxiv.org/abs/2203.07316}{{\ttfamily arXiv:2203.07316
  [hep-ex]}}.

\bibitem{MATHUSLA:2020uve}
{\bfseries MATHUSLA} Collaboration, C.~Alpigiani {\em et al.}, ``{An Update to
  the Letter of Intent for MATHUSLA: Search for Long-Lived Particles at the
  HL-LHC},'' \href{http://arxiv.org/abs/2009.01693}{{\ttfamily arXiv:2009.01693
  [physics.ins-det]}}.

\bibitem{Aberle:2839677}
{\bfseries SHiP} Collaboration, O.~Aberle {\em et al.}, ``{BDF/SHiP at the ECN3
  high-intensity beam facility},'' tech. rep., CERN, Geneva, 2022.
\newblock \url{http://cds.cern.ch/record/2839677}.

\bibitem{Altmannshofer:2009ma}
W.~Altmannshofer, A.~J. Buras, D.~M. Straub, and M.~Wick, ``{New strategies for
  New Physics search in $B \to K^{*} \nu \bar{\nu}$, $B \to K \nu \bar{\nu}$
  and $B \to X_{s} \nu \bar{\nu}$ decays},''
  \href{http://dx.doi.org/10.1088/1126-6708/2009/04/022}{{\em JHEP} {\bfseries
  04} (2009) 022}, \href{http://arxiv.org/abs/0902.0160}{{\ttfamily
  arXiv:0902.0160 [hep-ph]}}.

\bibitem{Okada:2019opp}
N.~Okada and D.~Raut, ``{Hunting inflatons at FASER},''
  \href{http://dx.doi.org/10.1103/PhysRevD.103.055022}{{\em Phys. Rev. D}
  {\bfseries 103} no.~5, (2021) 055022},
  \href{http://arxiv.org/abs/1910.09663}{{\ttfamily arXiv:1910.09663
  [hep-ph]}}.

\bibitem{Banerjee:2020kww}
A.~Banerjee, H.~Kim, O.~Matsedonskyi, G.~Perez, and M.~S. Safronova, ``{Probing
  the Relaxed Relaxion at the Luminosity and Precision Frontiers},''
  \href{http://dx.doi.org/10.1007/JHEP07(2020)153}{{\em JHEP} {\bfseries 07}
  (2020) 153}, \href{http://arxiv.org/abs/2004.02899}{{\ttfamily
  arXiv:2004.02899 [hep-ph]}}.

\bibitem{Csaki:2020zqz}
C.~Cs\'aki, R.~T. D'Agnolo, M.~Geller, and A.~Ismail, ``{Crunching Dilaton,
  Hidden Naturalness},''
  \href{http://dx.doi.org/10.1103/PhysRevLett.126.091801}{{\em Phys. Rev.
  Lett.} {\bfseries 126} (2021) 091801},
  \href{http://arxiv.org/abs/2007.14396}{{\ttfamily arXiv:2007.14396
  [hep-ph]}}.

\bibitem{Li:2022zgr}
S.~Li, H.~Song, S.~Su, and W.~Su, ``{Light Scalars at FASER},''
  \href{http://arxiv.org/abs/2212.06186}{{\ttfamily arXiv:2212.06186
  [hep-ph]}}.

\bibitem{Batell:2018fqo}
B.~Batell, A.~Freitas, A.~Ismail, and D.~Mckeen, ``{Probing Light Dark Matter
  with a Hadrophilic Scalar Mediator},''
  \href{http://dx.doi.org/10.1103/PhysRevD.100.095020}{{\em Phys. Rev. D}
  {\bfseries 100} no.~9, (2019) 095020},
  \href{http://arxiv.org/abs/1812.05103}{{\ttfamily arXiv:1812.05103
  [hep-ph]}}.

\bibitem{Demidov:2022ijc}
S.~Demidov, D.~Gorbunov, and D.~Kalashnikov, ``{Sgoldstino signal at FASER:
  prospects in searches for supersymmetry},''
  \href{http://dx.doi.org/10.1007/JHEP08(2022)155}{{\em JHEP} {\bfseries 08}
  (2022) 155}, \href{http://arxiv.org/abs/2202.05190}{{\ttfamily
  arXiv:2202.05190 [hep-ph]}}.

\bibitem{Berlin:2018jbm}
A.~Berlin and F.~Kling, ``{Inelastic Dark Matter at the LHC Lifetime Frontier:
  ATLAS, CMS, LHCb, CODEX-b, FASER, and MATHUSLA},''
  \href{http://dx.doi.org/10.1103/PhysRevD.99.015021}{{\em Phys. Rev. D}
  {\bfseries 99} no.~1, (2019) 015021},
  \href{http://arxiv.org/abs/1810.01879}{{\ttfamily arXiv:1810.01879
  [hep-ph]}}.

\bibitem{Dienes:2023uve}
K.~R. Dienes, J.~L. Feng, M.~Fieg, F.~Huang, S.~J. Lee, and B.~Thomas,
  ``{Extending the Discovery Potential for Inelastic-Dipole Dark Matter with
  FASER},'' \href{http://arxiv.org/abs/2301.05252}{{\ttfamily arXiv:2301.05252
  [hep-ph]}}.

\bibitem{Li:2021rzt}
J.~Li, T.~Nomura, and T.~Shimomura, ``{Inelastic dark matter from dark Higgs
  boson decays at FASER},''
  \href{http://dx.doi.org/10.1007/JHEP09(2022)140}{{\em JHEP} {\bfseries 09}
  (2022) 140}, \href{http://arxiv.org/abs/2112.12432}{{\ttfamily
  arXiv:2112.12432 [hep-ph]}}.

\bibitem{Bertuzzo:2022ozu}
E.~Bertuzzo, A.~Scaffidi, and M.~Taoso, ``{Searching for inelastic dark matter
  with future LHC experiments},''
  \href{http://dx.doi.org/10.1007/JHEP08(2022)100}{{\em JHEP} {\bfseries 08}
  (2022) 100}, \href{http://arxiv.org/abs/2201.12253}{{\ttfamily
  arXiv:2201.12253 [hep-ph]}}.

\bibitem{Batell:2021aja}
B.~Batell, J.~L. Feng, A.~Ismail, F.~Kling, R.~M. Abraham, and S.~Trojanowski,
  ``{Discovering dark matter at the LHC through its nuclear scattering in
  far-forward emulsion and liquid argon detectors},''
  \href{http://dx.doi.org/10.1103/PhysRevD.104.035036}{{\em Phys. Rev. D}
  {\bfseries 104} no.~3, (2021) 035036},
  \href{http://arxiv.org/abs/2107.00666}{{\ttfamily arXiv:2107.00666
  [hep-ph]}}.

\bibitem{Boyarsky:2021moj}
A.~Boyarsky, O.~Mikulenko, M.~Ovchynnikov, and L.~Shchutska, ``{Searches for
  new physics at SND@LHC},''
  \href{http://dx.doi.org/10.1007/JHEP03(2022)006}{{\em JHEP} {\bfseries 03}
  (2022) 006}, \href{http://arxiv.org/abs/2104.09688}{{\ttfamily
  arXiv:2104.09688 [hep-ph]}}.

\bibitem{Jodlowski:2020vhr}
K.~Jod\l{}owski and S.~Trojanowski, ``{Neutrino beam-dump experiment with FASER
  at the LHC},'' \href{http://dx.doi.org/10.1007/JHEP05(2021)191}{{\em JHEP}
  {\bfseries 05} (2021) 191}, \href{http://arxiv.org/abs/2011.04751}{{\ttfamily
  arXiv:2011.04751 [hep-ph]}}.

\bibitem{Ismail:2021dyp}
A.~Ismail, S.~Jana, and R.~M. Abraham, ``{Neutrino up-scattering via the dipole
  portal at forward LHC detectors},''
  \href{http://dx.doi.org/10.1103/PhysRevD.105.055008}{{\em Phys. Rev. D}
  {\bfseries 105} no.~5, (2022) 055008},
  \href{http://arxiv.org/abs/2109.05032}{{\ttfamily arXiv:2109.05032
  [hep-ph]}}.

\bibitem{Ovchynnikov:2023wgg}
M.~Ovchynnikov and J.-Y. Zhu, ``{Search for the dipole portal of heavy neutral
  leptons at future colliders},''
  \href{http://arxiv.org/abs/2301.08592}{{\ttfamily arXiv:2301.08592
  [hep-ph]}}.

\bibitem{Liu:2023nxi}
W.~Liu and Y.~Zhang, ``{Testing Neutrino Dipole Portal by Long-lived Particle
  Detectors at the LHC},'' \href{http://arxiv.org/abs/2302.02081}{{\ttfamily
  arXiv:2302.02081 [hep-ph]}}.

\bibitem{Cheung:2022kjd}
K.~Cheung and C.~J. Ouseph, ``{Sensitivities on dark photon from the forward
  physics experiments},'' \href{http://dx.doi.org/10.1007/JHEP10(2022)196}{{\em
  JHEP} {\bfseries 10} (2022) 196},
  \href{http://arxiv.org/abs/2208.04523}{{\ttfamily arXiv:2208.04523
  [hep-ph]}}.

\bibitem{Kling:2022ykt}
F.~Kling, J.-L. Kuo, S.~Trojanowski, and Y.-D. Tsai, ``{FLArE up dark sectors
  with EM form factors at the LHC forward physics facility},''
  \href{http://dx.doi.org/10.1016/j.nuclphysb.2023.116103}{{\em Nucl. Phys. B}
  {\bfseries 987} (2023) 116103},
  \href{http://arxiv.org/abs/2205.09137}{{\ttfamily arXiv:2205.09137
  [hep-ph]}}.

\bibitem{Kling:2022ehv}
F.~Kling and P.~Qu\'\i{}lez, ``{ALP searches at the LHC: FASER as a
  light-shining-through-walls experiment},''
  \href{http://dx.doi.org/10.1103/PhysRevD.106.055036}{{\em Phys. Rev. D}
  {\bfseries 106} no.~5, (2022) 055036},
  \href{http://arxiv.org/abs/2204.03599}{{\ttfamily arXiv:2204.03599
  [hep-ph]}}.

\bibitem{Jodlowski:2019ycu}
K.~Jod\l{}owski, F.~Kling, L.~Roszkowski, and S.~Trojanowski, ``{Extending the
  reach of FASER, MATHUSLA, and SHiP towards smaller lifetimes using secondary
  particle production},''
  \href{http://dx.doi.org/10.1103/PhysRevD.101.095020}{{\em Phys. Rev. D}
  {\bfseries 101} no.~9, (2020) 095020},
  \href{http://arxiv.org/abs/1911.11346}{{\ttfamily arXiv:1911.11346
  [hep-ph]}}.

\bibitem{Bakhti:2020szu}
P.~Bakhti, Y.~Farzan, and S.~Pascoli, ``{Discovery potential of FASER$\nu$ with
  contained vertex and through-going events},''
  \href{http://dx.doi.org/10.1007/JHEP04(2021)075}{{\em JHEP} {\bfseries 04}
  (2021) 075}, \href{http://arxiv.org/abs/2010.16312}{{\ttfamily
  arXiv:2010.16312 [hep-ph]}}.

\bibitem{Ansarifard:2021elw}
S.~Ansarifard and Y.~Farzan, ``{Neutral exotica at FASER\ensuremath{\nu} and
  SND@LHC},'' \href{http://dx.doi.org/10.1007/JHEP02(2022)049}{{\em JHEP}
  {\bfseries 02} (2022) 049}, \href{http://arxiv.org/abs/2109.13962}{{\ttfamily
  arXiv:2109.13962 [hep-ph]}}.

\bibitem{Cerci:2021nlb}
S.~Cerci {\em et al.}, ``{FACET: A new long-lived particle detector in the very
  forward region of the CMS experiment},''
  \href{http://dx.doi.org/10.1007/JHEP06(2022)110}{{\em JHEP} {\bfseries 2022}
  no.~06, (2022) 110}, \href{http://arxiv.org/abs/2201.00019}{{\ttfamily
  arXiv:2201.00019 [hep-ex]}}.

\bibitem{Helo:2018qej}
J.~C. Helo, M.~Hirsch, and Z.~S. Wang, ``{Heavy neutral fermions at the
  high-luminosity LHC},'' \href{http://dx.doi.org/10.1007/JHEP07(2018)056}{{\em
  JHEP} {\bfseries 07} (2018) 056},
  \href{http://arxiv.org/abs/1803.02212}{{\ttfamily arXiv:1803.02212
  [hep-ph]}}.

\bibitem{DeVries:2020jbs}
J.~De~Vries, H.~K. Dreiner, J.~Y. G\"unther, Z.~S. Wang, and G.~Zhou,
  ``{Long-lived Sterile Neutrinos at the LHC in Effective Field Theory},''
  \href{http://dx.doi.org/10.1007/JHEP03(2021)148}{{\em JHEP} {\bfseries 03}
  (2021) 148}, \href{http://arxiv.org/abs/2010.07305}{{\ttfamily
  arXiv:2010.07305 [hep-ph]}}.

\bibitem{Cottin:2021lzz}
G.~Cottin, J.~C. Helo, M.~Hirsch, A.~Titov, and Z.~S. Wang, ``{Heavy neutral
  leptons in effective field theory and the high-luminosity LHC},''
  \href{http://dx.doi.org/10.1007/JHEP09(2021)039}{{\em JHEP} {\bfseries 09}
  (2021) 039}, \href{http://arxiv.org/abs/2105.13851}{{\ttfamily
  arXiv:2105.13851 [hep-ph]}}.

\bibitem{Dercks:2018eua}
D.~Dercks, J.~De~Vries, H.~K. Dreiner, and Z.~S. Wang, ``{R-parity Violation
  and Light Neutralinos at CODEX-b, FASER, and MATHUSLA},''
  \href{http://dx.doi.org/10.1103/PhysRevD.99.055039}{{\em Phys. Rev. D}
  {\bfseries 99} no.~5, (2019) 055039},
  \href{http://arxiv.org/abs/1810.03617}{{\ttfamily arXiv:1810.03617
  [hep-ph]}}.

\bibitem{Dreiner:2022swd}
H.~K. Dreiner, D.~K\"ohler, S.~Nangia, and Z.~S. Wang, ``{Searching for a
  single photon from lightest neutralino decays in R-parity-violating
  supersymmetry at FASER},''
  \href{http://dx.doi.org/10.1007/JHEP02(2023)120}{{\em JHEP} {\bfseries 02}
  (2023) 120}, \href{http://arxiv.org/abs/2207.05100}{{\ttfamily
  arXiv:2207.05100 [hep-ph]}}.

\bibitem{Darme:2020ral}
L.~Darm\'e, S.~A.~R. Ellis, and T.~You, ``{Light Dark Sectors through the
  Fermion Portal},'' \href{http://dx.doi.org/10.1007/JHEP07(2020)053}{{\em
  JHEP} {\bfseries 07} (2020) 053},
  \href{http://arxiv.org/abs/2001.01490}{{\ttfamily arXiv:2001.01490
  [hep-ph]}}.

\bibitem{Araki:2022xqp}
T.~Araki, K.~Asai, H.~Otono, T.~Shimomura, and Y.~Takubo, ``{Search for Lepton
  Flavor Violating Decay at FASER},''
  \href{http://arxiv.org/abs/2210.12730}{{\ttfamily arXiv:2210.12730
  [hep-ph]}}.

\bibitem{Mao:2023zzk}
Y.-n. Mao, K.~Wang, and Z.~S. Wang, ``{Can we discover lepton number violation
  with LHC far detectors?},'' \href{http://arxiv.org/abs/2305.03908}{{\ttfamily
  arXiv:2305.03908 [hep-ph]}}.

\bibitem{Kling:2020mch}
F.~Kling and S.~Trojanowski, ``{Looking forward to test the KOTO anomaly with
  FASER},'' \href{http://dx.doi.org/10.1103/PhysRevD.102.015032}{{\em Phys.
  Rev. D} {\bfseries 102} no.~1, (2020) 015032},
  \href{http://arxiv.org/abs/2006.10630}{{\ttfamily arXiv:2006.10630
  [hep-ph]}}.

\bibitem{Carmona:2021seb}
A.~Carmona, C.~Scherb, and P.~Schwaller, ``{Charming ALPs},''
  \href{http://dx.doi.org/10.1007/JHEP08(2021)121}{{\em JHEP} {\bfseries 08}
  (2021) 121}, \href{http://arxiv.org/abs/2101.07803}{{\ttfamily
  arXiv:2101.07803 [hep-ph]}}.

\bibitem{Kling:2021gos}
F.~Kling and L.~J. Nevay, ``{Forward neutrino fluxes at the LHC},''
  \href{http://dx.doi.org/10.1103/PhysRevD.104.113008}{{\em Phys. Rev. D}
  {\bfseries 104} no.~11, (2021) 113008},
  \href{http://arxiv.org/abs/2105.08270}{{\ttfamily arXiv:2105.08270
  [hep-ph]}}.

\bibitem{Bahraminasr:2020ssz}
M.~Bahraminasr, P.~Bakhti, and M.~Rajaee, ``{Sensitivities to secret neutrino
  interaction at FASER\ensuremath{\nu}},''
  \href{http://dx.doi.org/10.1088/1361-6471/ac11c2}{{\em J. Phys. G} {\bfseries
  48} no.~9, (2021) 095001}, \href{http://arxiv.org/abs/2003.09985}{{\ttfamily
  arXiv:2003.09985 [hep-ph]}}.

\bibitem{Ansarifard:2021dju}
S.~Ansarifard and Y.~Farzan, ``{Excess of tau events at SND@LHC, FASER$\nu $
  and FASER$\nu $2},''
  \href{http://dx.doi.org/10.1140/epjc/s10052-022-10512-9}{{\em Eur. Phys. J.
  C} {\bfseries 82} no.~6, (2022) 568},
  \href{http://arxiv.org/abs/2112.08799}{{\ttfamily arXiv:2112.08799
  [hep-ph]}}.

\bibitem{Aloni:2022ebm}
D.~Aloni and A.~Dery, ``{Revisiting leptonic non-unitarity in light of
  FASER$\nu$},'' \href{http://arxiv.org/abs/2211.09638}{{\ttfamily
  arXiv:2211.09638 [hep-ph]}}.

\bibitem{Falkowski:2021bkq}
A.~Falkowski, M.~Gonz\'alez-Alonso, J.~Kopp, Y.~Soreq, and Z.~Tabrizi, ``{EFT
  at FASER\ensuremath{\nu}},''
  \href{http://dx.doi.org/10.1007/JHEP10(2021)086}{{\em JHEP} {\bfseries 10}
  (2021) 086}, \href{http://arxiv.org/abs/2105.12136}{{\ttfamily
  arXiv:2105.12136 [hep-ph]}}.

\bibitem{Cheung:2021tmx}
K.~Cheung, C.~J. Ouseph, and T.~Wang, ``{Non-standard neutrino and Z'
  interactions at the FASER\ensuremath{\nu} and the LHC},''
  \href{http://dx.doi.org/10.1007/JHEP12(2021)209}{{\em JHEP} {\bfseries 12}
  (2021) 209}, \href{http://arxiv.org/abs/2111.08375}{{\ttfamily
  arXiv:2111.08375 [hep-ph]}}.

\bibitem{Cheung:2023gwm}
K.~Cheung, T.~T.~Q. Nguyen, and C.~J. Ouseph, ``{Leptoquark Search at the
  Forward Physics Facility},''
  \href{http://arxiv.org/abs/2302.05461}{{\ttfamily arXiv:2302.05461
  [hep-ph]}}.

\bibitem{Cheung:2022oji}
K.~Cheung and C.~J. Ouseph, ``{Constraining the Active-to-Heavy-Neutrino
  transitional magnetic moments associated with the $Z'$ interactions at
  FASER$\nu$},'' \href{http://arxiv.org/abs/2205.11077}{{\ttfamily
  arXiv:2205.11077 [hep-ph]}}.

\bibitem{Kelly:2021mcd}
K.~J. Kelly, F.~Kling, D.~Tuckler, and Y.~Zhang, ``{Probing neutrino-portal
  dark matter at the Forward Physics Facility},''
  \href{http://dx.doi.org/10.1103/PhysRevD.105.075026}{{\em Phys. Rev. D}
  {\bfseries 105} no.~7, (2022) 075026},
  \href{http://arxiv.org/abs/2111.05868}{{\ttfamily arXiv:2111.05868
  [hep-ph]}}.

\bibitem{Ismail:2020yqc}
A.~Ismail, R.~Mammen~Abraham, and F.~Kling, ``{Neutral current neutrino
  interactions at FASER$\nu$},''
  \href{http://dx.doi.org/10.1103/PhysRevD.103.056014}{{\em Phys. Rev. D}
  {\bfseries 103} no.~5, (2021) 056014},
  \href{http://arxiv.org/abs/2012.10500}{{\ttfamily arXiv:2012.10500
  [hep-ph]}}.

\bibitem{MammenAbraham:2023psg}
R.~Mammen~Abraham, S.~Foroughi-Abari, F.~Kling, and Y.-D. Tsai, ``{Neutrino
  Electromagnetic Properties and the Weak Mixing Angle at the LHC Forward
  Physics Facility},'' \href{http://arxiv.org/abs/2301.10254}{{\ttfamily
  arXiv:2301.10254 [hep-ph]}}.

\bibitem{Bai:2020ukz}
W.~Bai, M.~Diwan, M.~V. Garzelli, Y.~S. Jeong, and M.~H. Reno, ``{Far-forward
  neutrinos at the Large Hadron Collider},''
  \href{http://dx.doi.org/10.1007/JHEP06(2020)032}{{\em JHEP} {\bfseries 06}
  (2020) 032}, \href{http://arxiv.org/abs/2002.03012}{{\ttfamily
  arXiv:2002.03012 [hep-ph]}}.

\bibitem{Maciula:2022lzk}
R.~Maciula and A.~Szczurek, ``{Far-forward production of charm mesons and
  neutrinos at forward physics facilities at the LHC and the intrinsic charm in
  the proton},'' \href{http://dx.doi.org/10.1103/PhysRevD.107.034002}{{\em
  Phys. Rev. D} {\bfseries 107} no.~3, (2023) 034002},
  \href{http://arxiv.org/abs/2210.08890}{{\ttfamily arXiv:2210.08890
  [hep-ph]}}.

\bibitem{Brdar:2021hpy}
V.~Brdar, A.~de~Gouv\^ea, P.~A.~N. Machado, and R.~Plestid, ``{Resonances in
  \ensuremath{\nu}\textasciimacron{}e-e- scattering below a TeV},''
  \href{http://dx.doi.org/10.1103/PhysRevD.105.093004}{{\em Phys. Rev. D}
  {\bfseries 105} no.~9, (2022) 093004},
  \href{http://arxiv.org/abs/2112.03283}{{\ttfamily arXiv:2112.03283
  [hep-ph]}}.

\bibitem{Li:2021tsy}
J.~Li, J.~Pei, L.~Ran, and W.~Zhang, ``{The quirk signal at FASER and FASER
  2},'' \href{http://dx.doi.org/10.1007/JHEP12(2021)109}{{\em JHEP} {\bfseries
  12} (2021) 109}, \href{http://arxiv.org/abs/2108.06748}{{\ttfamily
  arXiv:2108.06748 [hep-ph]}}.

\bibitem{Muontalk}
S.~Trojanowski, ``High-energy muons at the Forward Physics Facility.''
  \href{https://indico.cern.ch/event/955956/contributions/4085699/attachments/2139583/3604681/muons_STrojanowski.pdf}{FPF
  Kickoff Meeting}, 2020.

\bibitem{Ariga:2023fjg}
A.~Ariga, R.~Balkin, I.~Galon, E.~Kajomovitz, and Y.~Soreq, ``{Hunting muonic
  forces at emulsion detectors},''
  \href{http://arxiv.org/abs/2305.03102}{{\ttfamily arXiv:2305.03102
  [hep-ph]}}.

\end{thebibliography}\endgroup

\end{document}